Title: Noncommutativity vs gauge symmetry
Authors: I.P. Zois (IHES and Oxford University)
Comments: 16 pages, tex, version accepted for publication
Report-no: MIOU-03-03
Journal-ref: MCFA Annals Vol 4 (2005) 198-208

In many aspects the most complicated foliated manifolds are those with 
nonvanishing Godbillon-Vey class. We argue that they probably do not appear in 
physics and that is due to gauge symmetry which prevents the foliation from 
becoming ``too wild''; that means that the foliation does not develop resilient
 leaves which, at least in codim-1, by Duminy's theorem are responsible for 
the nontriviality of the GV-class.

\documentstyle[11pt,amsfonts,amssymb,amscd]{article}
\begin{document}
\title{Noncommutativity vs gauge symmetry}
\author{Ioannis P. \ ZOIS\thanks{current address: School of Mathematics, Cardiff University, PO Box 926, Cardiff CF24 4YH, 
UK, e-mail: zoisip@cf.ac.uk}\\
\\
IHES, 35 route de Chartres,\\
 F-91440, Bures-sur-Yvette, France
\\
and
\\
Mathematical Institute, 24-29 St. Giles',\\
 Oxford OX1 3LB, England
\\}
\date{}
\maketitle
\begin{abstract}

In many aspects the most complicated foliated manifolds are those with 
nonvanishing Godbillon-Vey class. We argue that they probably do not appear in 
physics and that is due to gauge symmetry which prevents the foliation from 
becoming ``too wild''; that means that the foliation does not develop resilient
 leaves which, at least in codim-1, by Duminy's theorem are responsible for 
the nontriviality of the GV-class.



\end{abstract}

\section{Introduction and Motivation}
 
Modern geometry is divided into two big areas: \emph{differential} and \emph{algebraic} geometry. The former studies 
structures
called (algebraic) \emph{varieties}, the later studies structures called \emph{manifolds}. Both these structures are 
spaces
 which locally--but not necessarily globally--"look like" Euclidean spaces (using local coordinate charts); both 
varieties and manifolds share 
one common important property:
the space of coordinate functions form a \textsl{commutative} ring (varieties) or algebra (manifolds). During the last 
20 years
or so a new geometry appeared which unifies and generalises the above two: it is called \emph{noncommutative geometry} and 
it studies
structures whose coordinate functions form a \emph{noncommutative} ring or algebra. As its main creator,
the French mathematician Alain Connes says, the motivation came directly from
quantum physics (but also from Index and K-Theory): as it is very well known, the difference between classical and quantum 
physics lies on Planck's constant
which from the mathematical point of view measures precisely the failure of commutativity between position and momentum. We
should also add that one of the main advantages of noncommutative geometry is that it offers the rigorous mathematical 
framework to study fractals and chaotic 
dynamical systems (as far as measure theory, analysis and topology are concerned).\\

Despite their increased level of difficulty and complexity, \emph{noncommutative spaces} have already been proved very
 useful in physical applications, and their use is bound to 
increase during the years to come, an obvious fact if one only reflects back on the motivation for their construction to 
begin with; perhaps the 
first example historically of the appearence of noncommutative spaces in physical applications is 
the study of  
the Integral Quantum Hall Effect (IQHE in short) on square lattices in a 
uniform \textsl{irrational} magnetic field, primarily through the work 
of Jean Bellissard (see \cite{connes}). By a noncommutative space we mean a 
space whose algebra of coordinate functions (or its analogue) is 
noncommutative. The existence of the uniform magnetic field $B$ turns the Brilluin zone from a
commutative to a \emph{noncommutative} 2-torus $T^{2}_{\theta}$ with irrational slope $\theta $, where $\theta$ is directly
related to the uniform magnetic field $B$.
 Essentially the same ideas were
carried over to the M-Theory context in the classic article due to 
Connes-Douglas-Schwarz (CDS in short) \cite{CDS}, and subsequently by Seiberg and Witten). From the  CDS article we 
learnt that M-Theory admits additional compactifications onto 
noncommutative tori which are higher dimensional versions of the 
noncommutative 2-torus. In that M-Theory context the role of the uniform 
magnetic field which is responsible for noncommutativity was played by the 
\emph{constant} 3-form field $C$ of $D=11$ 
supergravity. One can study the noncommutative tori either 
\textsl{algebraically} 
using their corresponding  noncommutative algebras of coordinate functions or 
\textsl{geometrically} using the so-called \emph{Kronecker foliations} 
on the tori. However these 
noncommutative tori in certain aspects which will hopefully become clearer 
below, are among
the ``less noncommutative spaces'' available. We would like to see if other 
``more noncommutative'' spaces may play any role in physics. That amounts to 
investigating what may happen as long as noncommutativity is concerned if the 
magnetic field (or the 3-from $C$) is no longer uniform. In other words we 
shall try to give (at least a partial) answer to the following question:\\

\textsl{How much noncommutativity can we have in physics?}\\ 

In the present article we shall argue that 
\emph{very noncommutative spaces}, examples of which are foliations with nonvanishing 
GV-class, probably do not 
appear in physics due to gauge symmetry; but spaces with ``medium or little
noncommutativity'' may very well appear and have significant consequences. Our answer is perhaps not totally unexpected,
after all Planck's constant is indeed very small, hence we have little noncommutativity in nature. Our reasoning however
 will be different in this article.\\

\section{Foliations and why they play a key role in noncommutative geometry}

Here we recall some basic facts about foliated manifolds: let $M$  be an 
$m$-dim 
smooth, closed and oriented manifold. A codim-$q$ foliation $F$ on 
$M$ is given by a codim-$q$ \emph{integrable} subbundle $F$ of the tangent 
bundle $TM$ of $M$. ``Integrable'' means that the tangential vector fields to 
the foliation, namely the smooth sections of the vector bundle $F$ over $M$, 
form a \emph{Lie subalgebra} of the Lie algebra of smooth sections of $TM$. An 
equivalent  local definition is given by a nonsingular decomposable $q$-form 
$\omega$ on $M$ satisfying the integrability condition

\begin{eqnarray}
\omega\wedge d\omega =0
\label{k1}
\end{eqnarray}

What this does in effect is that it gives a decomposition of $M$ into a 
disjoint union of codim-$q$ (and hence of dimension $(m-q)$) immersed and 
connected submanifolds of $M$ called leaves. The tangent spaces over the
 leaves are defined precisely by the vector fields which are annihilated by 
$\omega $.
Basic examples of foliations are Cartesian products and the total space of fibre bundles (the fibres are 
the leaves). Two
 important differences between foliations and fibre bundles are:\\
1. the topology of each leaf may vary (some leaves may be compact but some 
others may not; the fundamental groups also vary) whereas in a fibre bundle 
all fibres ``look the same'' as the typical fibre.
\\
2. the relative geometry of neighbouring leaves is far more complicated 
because they may ``spiral'' over each other without intersecting.\\

It is known that 
\emph{foliated manifolds} (more precisely the spaces of leaves of  foliated 
manifolds) provide an excellent list of 
examples of noncommutative spaces (see \cite{connes}). Connes has given a 
recipe how to construct a corresponding algebra, more precisely a $C^{*}$-algebra, for every foliation using the 
\emph{holonomy groupoid} of the foliation. The holonomy groupoid conceptually 
is a suitable generalisation of the holonomy of the connection on a vector 
bundle.\\
 A 
noncommutative space is a space whose algebra of coordinate functions is 
noncommutative. Thus the corresponding algebra of a foliation can be thought 
of as the space of coordinate functions on its topologically ill-behaved space 
of leaves. We say
that foliations provide an excellent list of examples of noncommutative 
spaces because although not every noncommutative algebra can be realised
 as the corresponding algebra to a foliation, it is known that all three 
types of factors can occure as corresponding algebras to foliations: 
Type I occurs at Reeb foliations, Type II at Kronecker foliations and Type
III at foliations with \textsl{nonvanishing Godbillon-Vey class}. So it is 
fair to say that foliations constitute the ``back bone'' of 
noncommutative spaces. It is understood that the source of the 
\textsl{noncommutativity}
at the corresponding $C^*$-algebra level is the \emph{holonomy} of the foliation. This encodes all 
the information about the foliation: it contains information about the 
topology of the leaves (primarily their fundamental groups) but it also 
contains information  about their relative geometry. It is perhaps 
illuminating 
to compare with the holonomy of a connection around a loop on a vector bundle: 
as it is well known this depends on the homotopy class of the loop, thus we 
see the fundamental group playing a role; but it also depends on the 
connection, ie how we parallel transport vectors along neighbouring fibres, so 
it depends on the relative geometry of the fibres. In the foliation case we 
``parallel transport'' transversals by sliding them along loops on the leaves.
\\

The \emph{most noncommutative spaces} arise from foliations with 
\emph{non-vanishing GV-class.} We want to argue here that at least these most 
extreme cases probably
 do not appear in physics. The reason, as was more or less suspected in 
\cite{ZBH}, is that gauge symmetry ``tames'' the foliation and prevents it 
from becoming ``very wild''.  Let us emphasise here that we are talking 
about foliations with nonvanishing GV-class and not about Type III factors in 
general.\\

Now we shall briefly define the Godbillon-Vey (GV for short) class: our 
codim-$q$ foliation is given by a
 $q$-form $\omega$ satisfying the integrability condition. By Frobenius' 
theorem the integrability condition is equivalent to 

\begin{eqnarray}
d\omega =a\wedge\omega
\label{k2}
\end{eqnarray}

for another 1-form $a$. Then the GV-class is the real $(2q+1)$ de Rham 
cohomology class $a\wedge (da)^{q}$. The geometric interpretation 
of $a$ is the following: since $F$ is a codim-$q$ integrable subbundle of the 
tangent bundle $TM$ of $M$, then its \textsl{transverse bundle} $Q:=TM/F$ is of
dim $q$ and hence its $q$-th exterior power $\Lambda ^{q}Q$ is a line bundle.
Then $a$ is a (Bott) connection on $\Lambda ^{q}Q$ with curvature $da$.\\

For foliations one also has the important notion of 
\emph{topological entropy} (eg see \cite{zois} or \cite{cc}) which roughly is 
another measure 
of how \textsl{wild} (or ``how noncommutative'') the quotient space of leaves
is. Roughly speaking topological entropy contains information only about the 
relative geometry of the leaves and not about their topology. So it encodes 
\emph{some} of the information contained in the holonomy groupoid. A Corollary 
to a deep theorem 
due to Gerard Duminy relates topological entropy with the Godbillon-Vey class
for codim-1 foliations: if the GV-class is 
nonzero then the topological entropy is also nonzero or equivalently if the
topological entropy vanishes, then so does the GV-class. However if the 
GV-class vanishes, then the topological entropy may or may not vanish, one 
does not know. That means that somehow the \textsl{topological entropy} of a 
foliation is a \emph{more delicate} notion than the \textsl{GV-class}.\\

\section{A list of Foliations in increasing order of complexity}

We said that the most noncommutative spaces arise from foliations with 
nonvanishing GV-class. At the other extreme of the spectrum we have the 
\textsl{less noncommutative} spaces: fibre bundles (we ignore the completely trivial example
of Cartesian products). Fibre bundles have 
corresponding algebras which are noncommutative but they are
\textsl{strongly Morita equivalent to commutative algebras} and their topological 
entropy vanishes, as does their GV-class. Nevertheless fibre bundles may
have non trivial holonomy, for example for a principal $G$-bundle where $G$ is 
a Lie group, the holonomy groupoid is a subgroup of the Lie group $G$ itself.\\

 The next \textsl{more noncommutative} spaces 
are foliations defined by \emph{closed forms}. (The integrability condition is 
trivially satisfied by closed forms). As was exhibited in \cite{ZBH}, fibre 
bundles with compact base manifold 
constitute particular examples of foliations defined by closed forms. 
Foliations defined by closed forms have always vanishing GV-class but their 
topological entropy vanishes only in codim-1 case. Moreover in codim-1, 
foliations with trivial holonomy are homoeomorphic to foliations defined by 
closed forms but we cannot deduce that a foliation defined by a closed form 
has trivial holonomy.\\

The next more noncommutative spaces to foliations defined by closed forms 
are the \emph{taut} foliations. We shall 
define them in the codim-1 case: A codim-1 foliation $F$ on an $m$-dim 
manifold $M$ is called 
\textsl{(topologically) taut} if there exists an $S^{1}$ intersecting 
transversely all leaves. A codim-1 foliation is called 
\textsl{(geometrically) taut} if there exists a metric for which all leaves 
are minimal submanifolds (ie have mean curvature zero). It is in fact a 
theorem to prove that these two 
definitions are equivalent. These definitions can be generalised for 
foliations of codim greater than 1.\\

For taut foliations we have Rumler's criterion: for each 
taut $(m-1)$-dim foliation 
there exists an $(m-1)$-form which is \emph{$F$-closed} and transverse to $F$.
The second condition means that the form is nonsingular when restricited
 on every leaf. The condition that the form is $F$-closed \emph{weakens} the 
condition of it being closed: in general for a $p$-dim foliation $F$, a 
$p$-form $\eta$ is called $F$-closed if $d\eta =0$ whenever at least $p$ 
vectors are tangent to $F$.\\

Moreover from the Hilsum-Skandalis theorem (see \cite{ms}) we learn that the 
corresponding algebra of foliations with a complete transversal (a complete 
transversal is a transversal intersecting all leaves) simplifies drastically 
(for taut foliations the complete transversal is just $S^{1}$): it is 
Morita equivallent to the algebra of the restricition of the holonomy groupoid 
to only the transversal itself. Taut foliations may prove a key ingredient in defining a 
\emph{Noncommutative Floer Homology}
for closed, oriented and connected 3-manifolds which are not necessarily homology 3-spheres (see \cite{ipzf}).\\ 

 Now the noncommutative spaces (noncommutative torus $T^{2}_{\theta}$) 
appearing in 
physics literature both in the Integral Quantum Hall Effect as well as in the 
CDS article are foliations defined by closed forms, in fact 
constant forms.
Their corresponding algebras are Type II 
since they are essentially algebras associated to \textsl{Kronecker 
foliations} on the torus defined by \emph{constant} differential forms. In the 
IQHE we have
a codim-1 foliation on a 2-torus defined by a constant 1-form (the ``slope''
$\theta $ of the Kronecker flow) which is 
essentially determined by the uniform magnetic field. Even the 
\textsl{noncommutative 2-torus} $T^{2}_{\theta }$ which has irrational slope 
$\theta $ is in fact 
\emph{homotopic} to the \textsl{commutative one} $T^2$ (usual torus or with 
rational slope $\theta $), thus having 
cyclic (co)homology and K-Theory \emph{isomorphic} to the de Rham cohomology 
and K-Theory respectively of the ordinary (commutative) 2-torus and hence from 
the point of view of \textsl{noncommutative topology} it is a rather trivial 
example. The only difference between the K-Theories of $T^{2}_{\theta }$ and 
$T^2$ is the \emph{order} of the Abelian groups.\\

Similar things hold for the CDS 
article: there one has the D=11 supergravity real 3-form potential $C$ which is also 
assumed to be \emph{constant} (and hence again closed) and that gives rise to 
higher dimensional Kronecker foliations on the compactified tori.\\

Now the \textsl{topological entropy of the Kronecker foliation} is  
\emph{zero}, so the foliations appearing in physics literature up to now are 
the most trivial noncommutative spaces. And 
since they are defined by constant forms (which are therefore closed), they
 also have vanishing GV-class.\\

It was the notion of topological entropy of foliations and its possible 
relation with physical entropy which served as our 
motivation for \cite{ZBH}: it is known that string theory gives in some cases 
an explanation of the microscopic origin of the black hole entropy. The 
compactified dimensions play an important role in this argument due to 
Horowitz, Strominger and Vafa back in 1996 (see \cite{vafa}). From the CDS
article we had some important new input, that M-theory admits additional 
compactifications to (foliated) noncommutative tori. So it is interesting
 to see what will happen if we assume that the \textsl{compactified 
dimensions} form not simply a noncommutative torus 
but a noncommutative torus with \emph{nonvanishing topological entropy}.  That 
might imply some modification
to the Beckenstein-Hawking area entropy formula for black holes (that's by the
string theory origin of black hole entropy). To guarantee
that the foliation has nonvanishing topological entropy, by Duminy's theorem,
one may assume that the foliation has nonvanishing GV-class (this is a 
sufficient but not necessary condition as we explained above). A first answer 
to this question based on \cite{z} which still needs further improvement was 
given in \cite{ZBH}.\\

\section{Foliations with nonvanishing GV-class vs gauge symmetry}

The starting point is to try to see if foliations with nonvanishing 
GV-class can occure as D=11 supergravity solutions following the D=11 supergravity 
interpretation in the CDS article.\\

Let us recall the bosonic part of the D=11 supergravity Lagrangian density
(we follow \cite{duff}):

\begin{eqnarray}
L_{11}=\frac{1}{2k_{11}^{2}}(R- \frac{1}{2.4!}G\wedge *G)-\frac{1}
{12k_{11}^{2}}.\frac{1}{3!4!^{2}}C\wedge G\wedge G + ``fermions''
\label{g1}
\end{eqnarray}

where by definition $G:=dC$. Our bosonic fields are the metric $g_{MN}$
with scalar curvature $R$
(all capital letters appearing as subscripts take the values $0,1,...,10$)
whose field 
equations are analogous to Einstein's equations with electromagnetic field 
in D=11: 

\begin{eqnarray}
R_{MN}-\frac{1}{2}g_{MN}R=\frac{1}{12}(G_{MSPQ}G_{N}^{SPQ}-\frac{1}{8}g_{MN}
G_{STXY}G^{STXY})
\label{g2}
\end{eqnarray}

and the real 3-form $C$ whose equations of motion are:

\begin{eqnarray}
d*G+\frac{1}{2}G\wedge G=0
\label{g3}
\end{eqnarray}

where $''*''$ denotes the Hodge dual.
For simplicity we set all fermionic fields equal to zero.
If now we assume that $C$ defines a codim-3 foliation namely $C\wedge G=0$,
that means that 
the Chern-Simons term in D=11 supergravity action $C\wedge G\wedge G$ vanishes and 
hence we are left with the equation of motion for $C$:

\begin{eqnarray}
d*dC=0
\label{g4}
\end{eqnarray}

The key question then is if such solutions 
exist, namely we want to see if there exist real 3-forms $C$ satisfying
the Euler-Lagrange equation \ref{g4} above
but at the same time they define a codim-3 foliation with nonvanishing 
GV-class, namely $C$ must also satisfy

\begin{eqnarray}
G:=dC=\theta \wedge C
\label{g5}
\end{eqnarray}

The above equation \ref{g5} is equivalent to $C\wedge G=0$ by the Frobenius theorem.
The Hodge star in equation \ref{g4} referes to some metric which satisfies 
Einstein's equations. The 1-form 
$\theta $ is the 1-form which appears in the definition of the 
corresponding GV-class which in this case will be 
$\theta \wedge (d\theta )^{3}$. [This 1-form $\theta $ can be seen as a (partial) Bott connection 
on the transverse bundle]. 
If we assume that the ambient 11-manifold is
closed, namely compact without boundary, then the equation of motion for $C$ 
is equivalent to $C$ being closed; hence the GV-class will vanish (we assume 
that $C$ is nonzero). In order 
then to have some hope to find a solution of the equations of motion which 
also define a codim-3 foliation with nonvanishing GV-class, we should either 
add a boundary or go to the noncompact case.\\ 

We can simplify our discussion further: since D=11 supergravity is very similar to 
gravity coupled to electromagnetism in odd dimensions, we start with the 
simplest case: gravity coupled to electromagnetism in dimension 3. In this 
case we denote by $A$ the electromagnetic potential and its field strength is 
denoted $F:=dA$. We have two options: either that the foliation is defined by 
the potential $A$ or by its field strength $F$. We start from the first: 
Maxwell's equation reads:

\begin{eqnarray}
d*dA=0
\label{g4'}
\end{eqnarray}

Since we want $A$ to define a codim-1 foliation it also has to satisfy

\begin{eqnarray}
A\wedge dA=0
\label{g5'}
\end{eqnarray}

where $F:=dA=\theta \wedge A$ for another 1-form $\theta $, thus the 
electromagnetic 
potential $A$ defines a codim-1 
foliation with nonvanishing GV-class $\theta \wedge d\theta $. As it is 
well-known by the 
Frobenius theorem the equation $A\wedge dA=0$ is equivalent to 
$dA=\theta \wedge A$.\\

Finally the metric has to satisfy Einstein's equations coupled to 
electromagnetism:

\begin{eqnarray}
R_{\mu \nu}-\frac{1}{2}g_{\mu \nu}R=\frac{1}{12}(F_{\mu \rho}F_{\nu}^{\rho}
-\frac{1}{8}g_{\mu \nu}F_{\sigma \rho}F^{\sigma \rho})
\label{g6}
\end{eqnarray}

Again on a closed 3-manifold Maxwell's equation is equivalent to $A$ being 
closed hence the GV-class will vanish (we assume $A$ is nonsingular). In order
 to hope to have nonvanishing 
GV-class (ie $A$ not closed) one has either to add a boundary or consider 
noncompact 3-manifolds.\\

Now here is the \textsl{key observation:} equation \ref{g5'} which says that the 
potential $A$
defines a codim-1 foliation with nonvanishing GV-class is \emph{not gauge 
invariant}.
Physics is not sensitive in gauge transformations, namely we can replace
$A$ by $\tilde{A}:=A+d\phi $ where $\phi $ is a zero form. A little 
calculation shows that although we started with a foliation defined by $A$, 
namely $A\wedge dA=0$, its gauge transform 
$\tilde{A}\wedge d\tilde{A}=A\wedge dA+d\phi \wedge dA=d\phi \wedge dA \neq 0$ is not zero in general. 
Hence the foliation structure can be 
completely destroyed by a gauge transformation!\\ 

{\bf Aside Note:} The remaining  term $d\phi \wedge dA$ 
however is a total derivative which will not contribute to 
the action if the manifold has no boundary; can we perhaps do something here
to save the day? We do not have a concrete suggestion but the following might 
be of relevance: as it is well-known $A_{\infty}$-algebras 
appear in the BV-formalism when the commutator of two BRST transformations 
does not close on shell. Some more evidence which made us think of 
$A_{\infty}$-algebras was from \cite{cs} where homotopy 
associative algebras appear in open string theory when the symplectic 
structure is lost, ie when 
the gauge invariant combination $\Omega :=B+F$ is no longer closed; this 
``looks similar''
to a codim-2 foliation defined by a 2-form $\Omega $ which is not closed, 
thus may have nonvanishing GV-class; physically this
corresponds to a curved D-brane embedded in a curved background. More 
precisely in \cite{cs} 
the authors investigate the deformation of D-brane world-volumes in curved 
backgrounds. They calculate the leading corrections to the boundary conformal 
field theory involving the background fields, and in particular they study 
the correlation functions of the resulting system. This allowed them to 
obtain the world-volume deformation, identifying the open string metric and 
the noncommutative deformation parameter. The picture that unfolded was the 
following: when the gauge invariant combination $\Omega = B + F$ is constant 
one obtains the standard Moyal deformation of the brane world-volume. 
Similarly, when $d\Omega = 0$ one obtains the noncommutative Kontsevich 
deformation, physically corresponding to a curved brane in a flat background. 
When the background is curved, $d\Omega \not= 0$, they find that the relevant 
algebraic structure is still based on the Kontsevich expansion, which now 
defines a nonassociative star product with an $A_{\infty}$ homotopy 
associative algebraic structure. They then recovered, within this formalism, 
some known results of Matrix theory in curved backgrounds.\\ 

Foliations are only invariant 
under multiplications of $A$ by a nowhere vanishing function $f$ 
so it seems that gauge invariance forbids 
foliations with nonvanishing GV-class to play any role in physics
even if they can exist as solutions of the equations of motion.
Thus we see that
foliations with nonvanishing GV-class are very \textsl{delicate} objects.\\
 
However if the 
foliation is defined by a \emph{closed} 1-form $A$, ie $dA=0$ 
(which is also an electromagnetic
potential), thus having vanishing GV-class,
then if we gauge transform $A$ to $\tilde{A}=A+d\phi $, then since 
$\tilde{A}$ is also closed
the foliation structure remains since $\tilde{A}\wedge d\tilde{A}=0$ and the 
new foliation defined by the closed 1-form $\tilde{A}$ has again vanishing 
GV-class. Hence 
foliations defined by closed forms (and consequently have 
vanishing GV-class) are very \emph{rigid} structures with respect to gauge 
transformations.\\ 

So to sum-up: If we start from a foliation defined by the electromagnetic 
potential $A$ which is a closed 1-form and thus has vanishing GV-class and we 
gauge transform $A$, we get another foliation defined by another closed 1-form
and thus has vanishing GV-class too; so in this case a gauge transformation 
does not change the GV-class. Yet if we start from a foliation with 
nonvanishing GV-class and we gauge transform it, this transformation will not 
only change the GV-class but it may destroy the foliation 
structure completely. This very different behaviour under gauge 
transformations between foliations 
defined by closed forms (which form a particular family of foliations with 
vanishing GV-class) and foliations with nonvanishing GV-class 
was surprising. Note that in codim-1 case a foliation 
defined by a closed 1-form has zero topological entropy as well.\\

There seems to be another alternative however, namely to assume that the 
foliation is defined by the field 
$F:=dA$ (or its dual), in this case we shall have a codim-2 foliation 
(provided $F$ is decomposable and nonsingular). But now $F$ is closed due 
to Bianchi identity, hence the codim-2 foliation now will have again vanishing 
GV-class. (Equivalently the derivative of 
the dual field vanishes due to Maxwell's equations). This picture is 
consistent with IQHE where the slope of 
the noncommutative 2-torus comes from a uniform magnetic field in 
$z$-coordinate (this is a component of $F$, not $A$). However the important 
difference here 
is that a \textsl{codim-2 foliation defined by a closed 2-form, although it 
will have zero GV-class}, it may have \emph{nonzero topological entropy.} 
Hence we might have
interesting phenomena appearing even when the Bianchi identity holds (which 
would mean vanishing GV-class). The problem is that we do not have a way to 
detect the appearance of the topological entropy in codimensions greater 
than 1. Duminy's criterion which uses the GV-class, although not absolutely 
satisfactory since it is 
not an if and only if statement, applies only to the codim-1 case.\\

There is also yet another setting, that of Seiberg-Witten (or monopole) 
equations in N=2 SUSY Yang-Mills theory as modified by Kronheimer and Mrowka 
(see \cite{Kr}) for a \emph{4-manifold with boundary.} The boundary 3-manifold 
may have a contact structure. Contact structures are ``cousins'' to 
foliations and moreover taut foliations correspond to tight contact 
structures. This possibility needs further study to see if one can get 
codim-1 foliations with nonvanishing GV-class in that set-up. One of the 
interesting points 
in that article is a correspondence between tight contact structures and taut 
foliations on the boundary and \textsl{symplectic structures} on the bulk.\\

\section{Remarks:}

Let us make some remarks:

{\bf 1.} The existence of the GV-class for a codim-$q$ foliation $q\geq 1$ 
follows from Bott's
vanishing theorem for the Pontrjagin classes in degree $k>2q$ of the
\emph{transverse bundle} of the foliation. One roughly can think of the 
GV-class as 
something like the ``corresponding (Abelian) Chern-Simons''
form in the following way: Bott's theorem says that given a smooth closed 
$m$-manifold $M$ with tangent bundle $TM$, if a codim-$q$ subbundle 
$F$ of $TM$ is integrable then the Pontrjagin classes of the transverse bundle 
$Q:=TM/F$ in degree $k>2q$ must vanish. Hence supposing $F$ defines indeed a 
codim-$q$ foliation, the first vanishing Pontrjagin 
class of its transverse bundle will be in degree $(2q+2)$. So the GV-class which 
is a real $(2q+1)$-form can be thought of as an ``Abelian Chern-Simons'' form 
whose exterior derivative will give the Pontrjagin class in degree $(2q+2)$.
This however vanishes by Bott's theorem and so the GV-class is indeed a 
cohomology class (ie it is closed).\\

However in this picture, strictly speaking, the GV-class really referes
to the transverse bundle and not to the foliation itself. Then the question
is: \textsl{up to what extend is the behaviour of the foliation determined by 
its transverse bundle?}\\ 

The GV-class is not the most natural object to study in order to deduce results
about the foliation itself because primarily it is some information
about the transverse bundle. This is so because as it is discussed in
 \cite{bott}, given a 
smooth closed $m$-manifold $M$, the
functor  from codim-$q$ Haefliger structures on $M$ (foliations are particular 
examples of Haefliger structures) to $GL(q; {\bf R})$-bundles over 
$M$ which
assignes the transverse bundle to any codim-$q$ Haefliger structure
is essentialy defined by the \emph{derivative} of the $\Gamma _{q}$-cocycle, 
the \textsl{Jacobian} of the local diffeomorphisms. This functor is not an 
equivalence of categories: it is neither
surjective (due to Bott's result not
every $GL(q;{\bf R})$-bundle can occure as the normal bundle of some codim-$q$ 
Haefliger structure since at least it has to satisfy Bott's theorem, ie it 
must have vanishing
Pontrjagin classes in degree $k>2q$). Nor is it injective since there are
different Haefliger structures with the same transverse bundle. So the lesson 
we have learnt is that \textsl{studying
foliations using secondary classes of their transverse bundle is like studying
functions by their 1st derivatives} and this is an approximation, one loses 
information.\\ 

The noncommutative geometry approach to study foliations is
probably a tool which is more ``sensitive'', hence it gives a better 
approximation; it is more delicate but more 
complicated because one uses the \emph{holonomy groupoid}
of the foliation itself, which is essentially the Haefliger cocycle
itself: one starts with the holonomy groupoid, then one takes the vector space 
of half-densities, equips it with a convolution product and with an involution
 and then
completes it to a $C^{*}$-algebra (or even more to a Hopf algebra) and
then one studies its cyclic cohomology. However passing from the holonomy 
groupoid to the $C^{*}$-algebra (even the reduced) amounts to loss of 
information again. At the K-Theory level this is probably not too bad
 since the 
Baum-Connes conjecture is true in many cases. Finally we get the 
\emph{transverse fundamental cyclic cocycle} (abreviated to \emph{``tfcc''}, 
see \cite{connes}) which for a codim-$q$ foliation belongs to the
$q$-th cyclic cohomology group of the corresponding algebra of the
foliation.  The reason why we believe that the noncommutative geometry 
approximation (Connes' approximation) is better than the Bott approximation is 
that at least in the codim-1 case the \textsl{derivative of the tfcc} 
\emph{is the GV-class} as explained in Connes' book. Hence the 
\textsl{tfcc} is a more \emph{delicate} 
object than the \textsl{GV-class}. That makes one 
to suspect that there may be a relation between the tfcc and the vanishing or 
not of the topological entropy of a foliation in 
a ``necessary and sufficient'' fashion which would improve considerably 
Duminy's result.\\ 

{\bf 2.} Our current level of understanding for 
\emph{codim-1 foliations} is the following (we assume compact manifolds): 
(fibre bundles) $\subset $ (foliations defined by \emph{closed} forms) 
$\subset $ (foliations with zero topological entropy) $\subset $ 
(foliations with zero GV-class). In higher codimensions one only has the 
following relations:
(fibre bundles) $\subset $ (foliations defined by \emph{closed} forms) 
$\subset $ (foliations with zero GV-class); the topological entropy may still
 be defined but we know nothing about when it is vanishing.
It would be desirable to introduce the tfcc 
into the picture as well.

So an important difference between the codim-1 and 
codim greater than 1 cases is the following: in codim-1 a foliation defined by 
a closed form 
has zero GV-class as well as topological entropy yet for codim greater than 1, 
foliations defined by closed forms have zero GV-class but may have non-zero 
topological entropy.\\

{\bf 3.} In general one can study foliations using two approaches: either 
using differential topology methods or operator algebraic tools. If one 
follows the first approach one meets
notions like the GV-class and topological entropy. The most complicated 
foliations using this language are those with nonvanishing GV-class. Probably 
they do not appear in physical applications. The second less complicated 
case is foliations with nonvanishing topological entropy. We would like to 
see if these appear in physical applications. Unfortunately we cannot see that
 clearly at this stage because we
do not have a satisfactory criterion which will indicate the existence of 
topological entropy. Duminy's theorem is a criterion only for the codim-1 case 
and even then not 
as satisfactory as one might wish since it is based on the GV-class in the 
way described above.\\

We would like to thank Alain Connes, T. Damour and A. Candel for useful 
discussions; moreover we wish to thank the IHES for its 
hospitality and for providing a stimulating atmosphere with excellent working 
conditions.\\

\end{document}